\documentclass[12pt]{article}
\usepackage{latexsym}
\usepackage{amsmath,amsfonts}
\usepackage{times}
\allowdisplaybreaks[4]

\catcode`\@=11

\newcount\hour
\newcount\minute
\newtoks\amorpm \hour=\time\divide\hour by 60\minute
=\time{\multiply\hour by 60 \global\advance\minute by-\hour}
\edef\standardtime{{\ifnum\hour<12 \global\amorpm={am}%
        \else\global\amorpm={pm}\advance\hour by-12 \fi
        \ifnum\hour=0 \hour=12 \fi
        \number\hour:\ifnum\minute<10
        0\fi\number\minute\the\amorpm}}
\edef\militarytime{\number\hour:\ifnum\minute<10 0\fi\number\minute}

\def\draftlabel#1{{\@bsphack\if@filesw {\let\thepage\relax
   \xdef\@gtempa{\write\@auxout{\string
      \newlabel{#1}{{\@currentlabel}{\thepage}}}}}\@gtempa
   \if@nobreak \ifvmode\nobreak\fi\fi\fi\@esphack}
        \gdef\@eqnlabel{#1}}
\def\@eqnlabel{}
\def\@vacuum{}
\def\marginnote#1{}
\def\draftmarginnote#1{\marginpar{\raggedright\scriptsize\tt#1}}
\def\draft{
        \pagestyle{plain}
        \overfullrule=2pt
        \oddsidemargin -.5truein
        \def\@oddhead{\sl \phantom{\today\quad\militarytime} \hfil
        \smash{\Large\sl DRAFT} \hfil \today\quad\militarytime}
        \let\@evenhead\@oddhead
        \let\label=\draftlabel
        \let\marginnote=\draftmarginnote
        \def\ps@empty{\let\@mkboth\@gobbletwo
        \def\@oddfoot{\hfil \smash{\Large\sl DRAFT} \hfil}
        \let\@evenfoot\@oddhead}
        \def\@eqnnum{(\theequation)\rlap{\kern\marginparsep\tt\@eqnlabel}%
        \global\let\@eqnlabel\@vacuum}  }

\newcommand{\rf}[1]{(\ref{#1})}
\renewcommand{\theequation}{\thesection.\arabic{equation}}
\renewcommand{\thefootnote}{\fnsymbol{footnote}}
\newcommand{\newsection}{   
\setcounter{equation}{0}\section}

\def\appendix#1{\addtocounter{section}{1}\setcounter{equation}{0}
\renewcommand{\thesection}{\Alph{section}}
\section*{Appendix \thesection\protect\indent \parbox[t]{11.15cm}{#1}}
\addcontentsline{toc}{section}{Appendix \thesection\ \ \ #1}}

\def\be{\begin{equation}}
\def\ee{\end{equation}}
\def\beq{\begin{eqnarray}}
\def\eeq{\end{eqnarray}}

\def\parline{\,\partial\kern -0.55em /\,\,}

\def\half{{\frac{1}{2}}}

\def\LL{{\cal L}}

\def\NN{{\cal N}}

\def\XX{{\cal X}}

\def\Nbf{{\bf N}}

\def\alphab{{\bar{\alpha}}}

\def\rhob{{\bar{\rho}}}
\def\etab{{\bar{\eta}}}

\def\Gb{\bar{G}}

\def\phik{|\phi\rangle}
\def\phibr{\langle\phi|}
\def\ck{|c\rangle}
\def\cbk{|\bar{c}\rangle}
\def\cbr{\langle c |}
\def\cbbr{\langle \bar{c}|}

\def\Phik{|\Phi\rangle}
\def\Phibr{\langle\Phi|}

\def\xik{|\xi\rangle}
\def\Xik{|\Xi\rangle}

\def\Ism{{\scriptscriptstyle I}}
\def\IIsm{{\scriptscriptstyle II}}
\def\FPsm{{\scriptscriptstyle FP}}

\def\smponetwo{{\scriptscriptstyle [1,2]}}

\def\alpar{\alpha\partial}
\def\albpar{{\bar\alpha\partial}}

\def\upsilonb{{\bar{\upsilon}}}

\def\Lb{\bar{L}}

\def\cb{\bar{c}}

\def\gb{\bar{g}}

\def\lb{\bar{l}}

\def\(A)dS{{\rm (A)dS}}

\def\ext{{\rm ext}}
\def\intrm{{\rm int}}
\def\gh{{\rm gh}}

\def\ssf{{\sf s}}
\def\ssfb{\bar{\sf s}}

\begin{document}


\begin{flushright}
FIAN-TD-2018-03 \qquad \ \ \ \ \ \ \ \ \ \ \  \\
arXiv: 1803.08421V2 [hep-th] \\
\end{flushright}

\vspace{1cm}

\begin{center}

{\Large \bf BRST-BV approach to continuous-spin field}

\vspace{2.5cm}

R.R. Metsaev%
\footnote{ E-mail: metsaev@lpi.ru
}

\vspace{1cm}

{\it Department of Theoretical Physics, P.N. Lebedev Physical
Institute, \\ Leninsky prospect 53,  Moscow 119991, Russia }

\vspace{3.5cm}

{\bf Abstract}

\end{center}

Using BRST-BV approach, massless and massive continuous-spin fields propagating in the flat space are studied. For such fields, BRST-BV gauge invariant Lagrangian is obtained. The Lagrangian and gauge transformations are constructed out of traceless gauge fields and traceless gauge transformation parameters. Interrelation between the BRST-BV Lagrangian and the Lagrangian for the continuous-spin fields in metric-like approach is demonstrated. Considering the BRST-BV Lagrangian in the Siegel gauge, we get gauge-fixed Lagrangian which is invariant under global BRST and antiBRST transformations.

\vspace{2cm}


\newpage
\renewcommand{\thefootnote}{\arabic{footnote}}
\setcounter{footnote}{0}

\newsection{ \large Introduction }

Continuous-spin field can be considered as a field theoretical realization of continuous-spin representation of the Poincar\'e algebra (see, e.g., Ref.\cite{Bekaert:2006py}).
In view of various interesting features, the continuous-spin field theory has attracted attention for a long period of time (see, e.g., Refs.\cite{Bekaert:2005in}-\cite{Alkalaev:2017hvj}).
For the review of recent developments in this topic see Ref.\cite{Bekaert:2017khg}. Perhaps one of the intriguing features of the continuous-spin field is that this field is decomposed into an infinite tower of coupled scalar, vector, and tensor fields which involves of every spin just once. We recall then that it is such tower of scalar, vector, and tensor fields that appears in the theory of higher-spin gauge field in AdS space \cite{Vasiliev:1990en}. Therefore we expect that certain intriguing interrelations between  the higher-spin gauge theory and the continuous-spin field theory might exist. Also it seems likely that some regimes in string theory can be related to the continuous-spin field theory (see, e.g., Refs.\cite{Savvidy:2003fx}). Regarding string theory, we note that all gauge invariant and Lorentz covariant formulations of string field theories are realized only in the framework of BRST-BV approach.%
\footnote{ Extended versions of the original BRST approach \cite{Becchi:1974xu}, which involve antifields \cite{Batalin:1981jr}, are referred to as BRST-BV approach in this paper.}
This is to say that BRST-BV approach turned out to be not only powerful method for the studying quantum properties of string field theories but also efficient method for  building classical string field theories \cite{Siegel:1984wx}.  Taking this into account it seems then worthwhile to apply a BRST-BV approach for the studying a continuous-spin field. This is what we do in this paper.

Lagrangian metric-like formulation of massless continuous-spin field in $R^{3,1}$ space-time was obtained in Ref.\cite{Schuster:2014hca}, while Lagrangian metric-like formulation of massless and massive continuous-spin fields in $R^{d-1,1}$ space-time with $d\geq 4$ was obtained in Ref.\cite{Metsaev:2016lhs}.
In this paper, our aim is twofold. First, we build BRST-BV Lagrangian formulation of massless and massive continuous-spin fields in $R^{d-1,1}$ space-time with $d\geq 4$. Second, we demonstrate the interrelation between our BRST-BV Lagrangian formulation and the metric-like Lagrangian formulation in Refs.\cite{Metsaev:2016lhs}.

The BRST--BV formulation of the continuous-spin fields we discuss in this paper
turns out to be closely related to the  metric-like formulation of the continuous-spin fields. The metric-like formulation of the  massless and massive continuous-spin fields in terms of double-traceless fields in $R^{d-1,1}$ space-time was obtained in  Refs.\cite{Metsaev:2016lhs}. In the framework of the BRST-BV formulation, we prefer however to deal with traceless fields. A metric-like formulation of the continuous-spin fields in terms of traceless fields can straightforwardly be obtained
from the metric-like formulation of the continuous-spin fields in terms of double-traceless fields. Therefore, for the reader convenience, in Sec.\ref{sect-02},
we briefly review the metric-like formulation of the continuous-spin fields in terms of the double-traceless fields developed in Ref.\cite{Metsaev:2016lhs} and then we use such
formulation for the derivation of the metric-like
formulation of the continuous-spin fields in terms of traceless fields. After that, in
Sec.\ref{sect-03}, we develop the BRST-BV formulation of
continuous-spin fields.  Namely, first, we begin with discussion of the field content
entering our BRST--BV formulation. Second, we present our result
for the gauge invariant BRST-BV Lagrangian for massless and massive continuous-spin fields. Third, we discuss the interrelation between the BRST-BV Lagrangian formulation and the metric-like Lagrangian formulation. Finally, using the Siegel gauge, we obtain gauge-fixed Lagrangian which is invariant under global BRST and antiBRST transformations.

\newsection{\large  Metric-like formulation of continuous-spin field in $R^{d-1,1}$  }\label{sect-02}

\noindent {\bf Lagrangian formulation in terms of double-traceless fields}. To discuss
metric-like Lagrangian formulation we use the following set of scalar, vector, and tensor fields of the $so(d-1,1)$ Lorentz algebra
\be \label{13032018-man01-01}
\phi^{a_1\ldots a_n}\,, \qquad n=0,1,2,\ldots, \infty\,.
\ee
In \rf{13032018-man01-01}, fields with $n=0$, $n=1$, and $n\geq 2$ are the respective scalar, vector, and rank-$n$ totally symmetric tensor fields of the $so(d-1,1)$ Lorentz algebra. Fields \rf{13032018-man01-01} with $n\geq 4$ are double-traceless, $\phi^{aabba_5\ldots a_n} =0$.

To streamline presentation of the Lagrangian formulation we introduce oscillators $\alpha^a$, $\upsilon$ to collect fields \rf{13032018-man01-01} into ket-vector $\phik$ defined by the relation
\be \label{13032018-man01-02}
\phik = \sum_{n=0}^\infty \frac{ \upsilon^n }{ n!\sqrt{n!} } \ \alpha^{a_1} \ldots \alpha^{a_n}  \phi^{a_1\ldots a_n}(x)|0\rangle\,.
\ee
Our convention for the oscillators may be found in the Appendix. Ket-vector \rf{13032018-man01-02} satisfies the following two algebraic constraints:
\be \label{13032018-man01-03}
(N_\alpha-N_\upsilon)\phik = 0 \,, \qquad (\alphab^2)^2\phik=0\,,
\ee
where the operators $N_\alpha$, $N_\upsilon$, $\alphab^2$ are defined in \rf{17032018-man01-04},\rf{17032018-man01-05} in the Appendix. The second constraint in \rf{13032018-man01-03} tells us that fields \rf{13032018-man01-01} with $n\geq 4$ are double-traceless.

In terms of the ket-vector $\phik$, action and Lagrangian take the form \cite{Metsaev:2016lhs},
\beq
\label{13032018-man01-04} && S = \int d^d x\, \LL\,,  \qquad \LL   = \half
\phibr (1-\frac{1}{4}\alpha^2\alphab^2)
(\Box - m^2)\phik + \half
\langle \Lb
\phi|
\Lb\phi\rangle \,,\qquad
\\
\label{13032018-man01-05} &&   \Lb \equiv \albpar - \half \alpar \bar\alpha^2 -
\Pi^\smponetwo \gb + \half g \bar\alpha^2\,,
\\
\label{13032018-man01-06} &&  g \equiv  g_\upsilon \upsilonb  \,,\qquad \qquad \gb \equiv -
\upsilon g_\upsilon\,, \qquad g_\upsilon \equiv \Bigl(\frac{1}{(N_\upsilon+1)(2N_\upsilon+d-2)}F_\upsilon\Bigr)^{1/2}\,, \qquad
\\
\label{13032018-man01-07} && F_\upsilon = \kappa^2 - N_\upsilon (N_\upsilon+d-3)m^2\,,
\eeq
where $\langle \phi|\equiv \phik^\dagger$, $|\Lb\phi\rangle\equiv \Lb\phik$, $\langle \Lb\phi| \equiv |\Lb\phi\rangle^\dagger$, while the definition of the operators $\Box$, $\alpar$, $\alpha^2$, $\Pi^\smponetwo$ is given in the Appendix (see \rf{17032018-man01-04}-\rf{17032018-man01-06}).
In \rf{13032018-man01-04},\rf{13032018-man01-07}, the $m$ is a mass parameter, while the $\kappa$ stands for a dimensionfull constant parameter. Requiring the action \rf{13032018-man01-04} to be real-valued (classical unitarity) and irreducible, we get the restrictions $\kappa^2>0$, $m^2 \leq  0$.

To discuss gauge symmetries of action \rf{13032018-man01-04} we introduce the following set of gauge transformation parameters:
\be \label{13032018-man01-08}
\xi^{a_1\ldots a_n}\,, \qquad n=0,1,2,\ldots, \infty\,.
\ee
In \rf{13032018-man01-08}, gauge transformation parameters with $n=0$, $n=1$, and $n\geq 2$ are the respective scalar, vector, and rank-$n$ totally symmetric tensor fields of the $so(d-1,1)$ Lorentz algebra. Gauge transformation parameters \rf{13032018-man01-08} with $n\geq 2$ are traceless, $\xi^{aaa_3\ldots a_n} =0$.

To simplify the presentation of gauge symmetries we collect gauge transformation parameters  \rf{13032018-man01-08} into ket-vector $\xik$ defined by the relation
\be \label{13032018-man01-09}
\xik = \sum_{n=0}^\infty \frac{ \upsilon^{n+1} }{ n!\sqrt{(n+1)!} } \ \alpha^{a_1} \ldots \alpha^{a_n}  \xi^{a_1\ldots a_n}(x)|0\rangle\,.
\ee
Ket-vector \rf{13032018-man01-09} satisfies the following two algebraic constraints:
\be \label{13032018-man01-10}
(N_\alpha-N_\upsilon + 1)\xik = 0 \,, \qquad \alphab^2\xik=0\,.
\ee
The second constraint in \rf{13032018-man01-10} tells us that $\xi^{a_1\ldots a_n}$ \rf{13032018-man01-08} with $n\geq 2$ are traceless.

By using ket-vectors $\phik$ \rf{13032018-man01-02} and $\xik$
\rf{13032018-man01-09}, we can present gauge transformations as
\be \label{13032018-man01-11}
\delta \phik = G \xik  \,,
\qquad
G \equiv
\alpar - g - \alpha^2 \frac{1}{2N_\alpha +d-2} \gb\,.
\ee

\noindent{\bf Lagrangian formulation in terms of traceless   fields}. Using the formulation in
terms of double-traceless fields discussed above, we can obtain a Lagrangian formulation in terms of
traceless fields in a rather straightforward way. To this end we decompose the double-traceless ket-vector $\phik$ \rf{13032018-man01-02} in two traceless
ket-vectors, denoted by $|\phi_\Ism \rangle$, $|\phi_\IIsm \rangle$,
\beq
\label{13032018-man01-14} && \phik = |\phi_\Ism \rangle   + \alpha^2 \NN |\phi_\IIsm \rangle\,,
\hspace{1.5cm} \NN\equiv \bigl((2N_\alpha+d)(2N_\alpha+d-2)\bigr)^{-1/2}\,,
\\
\label{13032018-man01-15}  && \alphab^2 |\phi_\Ism \rangle =0\,,  \qquad \alphab^2 |\phi_\IIsm \rangle =0\,.
\eeq
Relations \rf{13032018-man01-15} tell us that the ket-vectors
$|\phi_\Ism \rangle$, $|\phi_\IIsm \rangle$ are traceless. Plugging $\phik$ \rf{13032018-man01-14} into \rf{13032018-man01-04}, we get the desired
representation for the Lagrangian in terms of the traceless ket-vectors,
\be \label{13032018-man01-16}
\LL  =    \half \langle \phi_\Ism | (\Box - m^2)  |\phi_\Ism \rangle - \half \langle \phi_\IIsm | (\Box - m^2)  |\phi_\IIsm \rangle + \half \langle \Lb \phi|\Lb\phi\rangle \,.
\ee
The $|\Lb\phi\rangle$ appearing in \rf{13032018-man01-16} can also be expressed in terms of the traceless ket-vectors $|\phi_\Ism\rangle$, $|\phi_\IIsm\rangle$ by using \rf{13032018-man01-14},\rf{13032018-man01-15},
\beq
\label{13032018-man01-17}  \Lb\phik & = &  \Lb_\Ism |\phi_\Ism\rangle + L_\IIsm |\phi_\IIsm\rangle\,,
\\
\label{13032018-man01-18}  && \Lb_\Ism \equiv \albpar - \gb\,,
\\
\label{13032018-man01-19}  && L_\IIsm \equiv - A^a\partial^a f_\upsilon  + g f_\upsilon^{-1}\,,
\\
\label{13032018-man01-20}  && f_\upsilon \equiv \Bigl(\frac{2N_\upsilon + d - 6}{2N_\upsilon+d-4}\Bigr)^{1/2}\,,
\eeq
where the operator $A^a$ is defined in \rf{17032018-man01-06}. Operators $g$, $\gb$ appearing in \rf{13032018-man01-18},\rf{13032018-man01-19} are given in \rf{13032018-man01-06}. In terms of the ket-vectors $|\phi_\Ism\rangle$, $|\phi_\IIsm\rangle$,
gauge transformations \rf{13032018-man01-11} take the following form:
\beq
\label{18032018-man01-01}  && \delta |\phi_\Ism\rangle = G_\Ism \xik\,, \hspace{1.1cm} G_\Ism \equiv A^a\partial^a - g\,,
\\
\label{18032018-man01-02} && \delta |\phi_\IIsm\rangle = \Gb_\IIsm \xik\,, \qquad  \Gb_\IIsm \equiv f_\upsilon \albpar - f_\upsilon^{-1}\gb\,,
\eeq
where the $\xik$ is given in \rf{13032018-man01-09}. Note that, for the derivation of gauge transformations \rf{18032018-man01-01},\rf{18032018-man01-02} we use \rf{13032018-man01-11} and the inverse to relation in \rf{13032018-man01-14}
\beq
\label{18032018-man01-03} && |\phi_\Ism\rangle = \Pi^\smponetwo \phik\,,
\\
\label{18032018-man01-04} && |\phi_\IIsm\rangle = \half \Bigl(\frac{2N_\alpha +d-2}{2N_\alpha+d}\Bigr)^{1/2}
\alphab^2 \phik\,,
\eeq
where the operator $ \Pi^\smponetwo$ is given in \rf{17032018-man01-06}.

To summarize, our formulations in terms of traceless fields and double-traceless fields are completely
equivalent.

\newsection{\large  BRST-BV formulation of continuous-spin field in $R^{d-1,1}$  }\label{sect-03}

\noindent {\bf Field content}. To describe a field content entering our BRST-BV formulation of a continuous-spin field we introduce Grassmann coordinate $\theta$, Grassmann odd oscillators $\eta$, $\rho$, and Grassmann even oscillators $\alpha^a$ and $\upsilon$. The coordinate $\theta$ and the oscillators $\eta$, $\rho$, $\upsilon$ transform as scalars of the $so(d-1,1)$ Lorentz algebra, while the oscillators $\alpha^a$ transform as vector of the $so(d-1,1)$ Lorentz algebra. Using the coordinate $\theta$ and the oscillators we introduce a ket-vector $\Phik$ by the relation
\be \label{11032018-man01-01}
\Phik = \Phi(x,\theta,\alpha,\upsilon,\eta,\rho)|0\rangle\,,
\ee
where the argument $x$ in \rf{11032018-man01-01} stands for coordinates $x^a$ of the space-time $R^{d-1,1}$, while the argument $\alpha$ in \rf{11032018-man01-01} stands for the vector oscillators $\alpha^a$.
By definition, field $\Phi$ \rf{11032018-man01-01} is Grassmann even. Also, by definition, ket-vector $\Phik$ \rf{11032018-man01-01} satisfies the following two algebraic constraints:
\beq
\label{11032018-man01-02} && (N_\alpha + N_\eta + N_\rho - N_\upsilon)\Phik = 0 \,,
\\
\label{11032018-man01-03} && \alphab^2\Phik=0\,.
\eeq
Usual scalar, vector, and tensor fields depending on the space-time coordinates $x^a$ are obtained by expanding $\Phi$ \rf{11032018-man01-01} into the Grassmann coordinate $\theta$ and the oscillators $\alpha^a$, $\upsilon$, $\eta$, $\rho$. Using algebraic constraints \rf{11032018-man01-02},\rf{11032018-man01-03}, we now illustrate a catalogue of the scalar, vector, and tensor fields entering ket-vector $\Phik$ \rf{11032018-man01-01}. To this end we note that the Taylor series expansion of ket-vector $\Phik$  \rf{11032018-man01-01} into the Grassmann coordinate $\theta$ and the Grassmann-odd oscillators $\eta$, $\rho$ is given by
\beq
\label{11032018-man01-04} \Phik & = & \phik + \theta |\phi_*\rangle\,,
\\
\label{11032018-man01-05} && \phik = |\phi_\Ism\rangle + \rho |c\rangle + \eta |\cb\rangle + \rho \eta |\phi_\IIsm\rangle\,,
\\
\label{11032018-man01-06} && |\phi_*\rangle = |\phi_{\Ism *}\rangle + \rho |\cb_*\rangle + \eta |c_*\rangle + \rho \eta |\phi_{\IIsm *}\rangle\,,
\eeq
where ket-vectors appearing on the right hand side of relations in \rf{11032018-man01-05},\rf{11032018-man01-06} depend, besides the space-time coordinates $x^a$, only on the oscillators $\alpha^a$, $\upsilon$. Throughout this paper, ket-vector $\phik$ \rf{11032018-man01-05} is referred to as ket-vector of fields, while ket-vector $|\phi_*\rangle$ \rf{11032018-man01-06} is referred to as ket-vector of antifields. Plugging $\Phik$ \rf{11032018-man01-04} into \rf{11032018-man01-02},\rf{11032018-man01-03}, we see that algebraic constraints for the ket-vectors $\phik$, $|\phi_*\rangle$ take the same form as the ones for the ket-vector $\Phik$ in \rf{11032018-man01-02},\rf{11032018-man01-03},
\beq
\label{11032018-man01-02-a1} && (N_\alpha + N_\eta + N_\rho - N_\upsilon)\phik = 0 \,, \qquad  \ \alphab^2\phik=0\,,
\\
\label{11032018-man01-02-a2} && (N_\alpha + N_\eta + N_\rho - N_\upsilon)|\phi_*\rangle = 0 \,, \qquad  \alphab^2 |\phi_*\rangle = 0\,.
\eeq
Now taking into account the first constraint in \rf{11032018-man01-02-a1}, it is easy to see that the ket-vectors $|\phi_{\Ism,\IIsm}\rangle$, $\ck$, $\cbk$ \rf{11032018-man01-05} depend on the oscillators $\alpha^a$, $\upsilon$ in the following way:
\beq
\label{11032018-man01-07}  && |\phi_\Ism\rangle = \sum_{n=0}^\infty \frac{ \upsilon^n }{ n!\sqrt{n!} } \ \alpha^{a_1} \ldots \alpha^{a_n}  \phi_\Ism^{a_1\ldots a_n}(x)|0\rangle\,,
\\
\label{11032018-man01-08}  && \ck  \ = \ \sum_{n=0}^\infty \frac{ \upsilon^{n+1} }{ n!\sqrt{(n+1)!} } \ \alpha^{a_1} \ldots \alpha^{a_n}  c^{a_1\ldots a_n}(x)|0\rangle\,,
\\
\label{11032018-man01-09}  &&  \cbk \ = \ \sum_{n=0}^\infty \frac{ \upsilon^{n+1} }{ n!\sqrt{(n+1)!} } \ \alpha^{a_1} \ldots \alpha^{a_n}  \cb^{a_1\ldots a_n}(x)|0\rangle\,,
\\
\label{11032018-man01-10}  && |\phi_\IIsm\rangle  = \sum_{n=0}^\infty \frac{ \upsilon^{n+2} }{ n!\sqrt{(n+2)!} } \ \alpha^{a_1} \ldots \alpha^{a_n}  \phi_\IIsm^{a_1\ldots a_n}(x)|0\rangle\,.
\eeq
From \rf{11032018-man01-02-a1},\rf{11032018-man01-02-a2}, we see that the ket-vectors $\phik$, $\phi_*\rangle$ satisfy the same algebraic constraints. Therefore the Taylor series expansion of the ket-vectors $|\phi_{*\Ism,\IIsm}\rangle$, $|c_*\rangle$, $|\cb_*\rangle$ \rf{11032018-man01-06} in the oscillators $\alpha^a$, $\upsilon$ takes the
same form as the one in \rf{11032018-man01-07}-\rf{11032018-man01-10},
\beq
\label{11032018-man01-11} && |\phi_{*\Ism}\rangle = \sum_{n=0}^\infty \frac{ \upsilon^n }{ n!\sqrt{n!} } \ \alpha^{a_1} \ldots \alpha^{a_n}  \phi_{*\Ism}^{a_1\ldots a_n}(x)|0\rangle\,,
\\
\label{11032018-man01-12} && |c_*\rangle  \ = \ \sum_{n=0}^\infty \frac{ \upsilon^{n+1} }{ n!\sqrt{(n+1)!} } \ \alpha^{a_1} \ldots \alpha^{a_n}  c_*^{a_1\ldots a_n}(x)|0\rangle\,,
\\
\label{11032018-man01-14} &&  |\cb_*\rangle \ = \ \sum_{n=0}^\infty \frac{ \upsilon^{n+1} }{ n!\sqrt{(n+1)!} } \ \alpha^{a_1} \ldots \alpha^{a_n}  \cb_*^{a_1\ldots a_n}(x)|0\rangle\,,
\\
\label{11032018-man01-15} && |\phi_{*\IIsm}\rangle  = \sum_{n=0}^\infty \frac{ \upsilon^{n+2} }{ n!\sqrt{(n+2)!} } \ \alpha^{a_1} \ldots \alpha^{a_n}  \phi_{*\IIsm}^{a_1\ldots a_n}(x)|0\rangle\,.
\eeq

To summarize, relations \rf{11032018-man01-07}-\rf{11032018-man01-10} illustrate the field content entering the ket-vector $\phik$ \rf{11032018-man01-05}, while relations \rf{11032018-man01-11}-\rf{11032018-man01-15} illustrate the antifield content entering ket-vector $|\phi_*\rangle$ \rf{11032018-man01-06}.
(Anti)fields with the $n=0$, $n=1$, and $n\geq 2$ are the respective scalar, vector, and totally symmetric rank-$n$ tensor fields of the $so(d - 1,1)$ Lorentz algebra.
The second constraints in \rf{11032018-man01-02-a1},\rf{11032018-man01-02-a2} tell us that all the tensor (anti)fields are realized as traceless tensors of the $so(d - 1,1)$ Lorentz algebra.%
\footnote{ The use of the traceless (anti)fields for the BRST-BV formulation of conformal fields and massless AdS fields may be found in the respective Ref.\cite{Metsaev:2015yyv} and Ref.\cite{Metsaev:2015oza}. Traceless fields also turn out to be convenient for the studying AdS/CFT correspondence in the framework of BRST approach \cite{Metsaev:2014vda}. Discussion of various unconstrained and constrained BRST formulations may be found, e.g., in Refs.\cite{Buchbinder:2001bs}-\cite{Reshetnyak:2018fvd}.}

\noindent {\bf BRST-BV Lagrangian and gauge symmetries}. We use the version of the BRST-BV formulation  discussed in Ref.\cite{Siegel:1984wx}.%
\footnote{ Discussion of other interesting version of BRST formulation with application for the studying massless continuous-spin field in $R^{3,1}$ may be found in Ref.\cite{Bengtsson:2013vra}. The BRST action in Ref.\cite{Bengtsson:2013vra} consists, besides $\Phibr Q \Phik$-term, the additional contribution with Lagrange multiplier fields. Also, in Ref.\cite{Bengtsson:2013vra}, there is interesting conjecture about interrelation between the higher-spin field theory in Refs.\cite{Francia:2002aa} and some regime in the continuous-spin field theory.}
Let us briefly review the main ingredients of the formulation in Ref.\cite{Siegel:1984wx}.  The gauge-invariant BRST-BV action takes the following form:
\be \label{11032018-man01-16}
S = \int d^dx\, \LL\,, \hspace{1cm} \LL = \half \int d\theta \Phibr Q \Phik\,,
\ee
where the BRST operator $Q$ entering BRST-BV Lagrangian \rf{11032018-man01-16} admits the following representation:
\be \label{11032018-man01-17}
Q =  \theta (\Box - M^2) + M^{\eta a} \partial^a + M^\eta + \half M^{\eta\eta} \partial_\theta\,.
\ee
In \rf{11032018-man01-17}, $\Box=\partial^a\partial^a$ is the D'Alembert operator in $R^{d-1,1}$, while $\partial_\theta$ is the left derivative of the Grassmann coordinate, $\partial_\theta = \partial/\partial\theta$. Operators $M^2$, $M^{\eta a}$, $M^\eta$, $M^{\eta\eta}$ entering $Q$ \rf{11032018-man01-17} depend only on the oscillators and do not depend on the Grassmann coordinate $\theta$, the space-time coordinates $x^a$, and the derivatives $\partial_\theta$, $\partial^a$.  In this paper, the operators  $M^{\eta a}$, $M^\eta$, $M^{\eta\eta}$ are referred to as spin operators, while the operator $M^2$ is referred to as square mass operator.  The nilpotence equation $Q^2=0$ implies the following (anti)commutation relations for the $M^2$ and the spin operators,
\beq
\label{15032018-man01-01} && \{ M^{\eta a},M^{\eta b}\} = -  \eta^{ab} M^{\eta\eta}\,,
\\
\label{15032018-man01-02} && \{ M^\eta ,M^\eta \} =   M^2 M^{\eta\eta}\,,
\\
\label{15032018-man01-03} && [M^2,M^{\eta a}]= 0\,, \hspace{1cm} [M^2,M^\eta]= 0\,, \hspace{1.2cm} [M^2,M^{\eta\eta}]= 0\,,
\\
\label{15032018-man01-04} && \{ M^{\eta a},M^\eta \}= 0\,, \qquad [M^{\eta a},M^{\eta\eta} ] = 0\,, \qquad [M^\eta,M^{\eta\eta}]= 0\,.
\eeq

Action \rf{11032018-man01-16} is invariant under gauge transformations given by
\be \label{15032018-man01-05}
\delta \Phik = Q \Xik, \qquad  \Xik  = \Xi(x,\theta,\alpha,\upsilon,\eta,\rho)|0\rangle\,.
\ee
Gauge transformation parameter $\Xi$ \rf{15032018-man01-05} depends on the space-time coordinates $x^a$, the Grassmann coordinate  $\theta$, and the oscillators $\alpha^a$, $\upsilon$, $\eta$, $\rho$. By definition, $\Xi$ \rf{15032018-man01-05} is Grassmann odd. Also, by definition, the ket-vector $\Xik$ satisfies the following two algebraic constraints:
\beq
\label{15032018-man01-06} && (N_\alpha + N_\eta + N_\rho - N_\upsilon)\Xik = 0 \,,
\\
\label{15032018-man01-07} && \alphab^2\Xik=0\,.
\eeq
We see that constraints for the $\Xik$ \rf{15032018-man01-06},\rf{15032018-man01-07} take the same form as the ones for the $\Phik$ \rf{11032018-man01-02},\rf{11032018-man01-03}. Therefore our analysis of the constraints for the $\Phik$ is easily extended to the case of the $\Xik$. Namely,  the $\Xik$ is built in terms of gauge transformation parameters which are scalar, vector, and totally symmetric traceless tensor fields of the $so(d-1,1)$ Lorentz algebra. Also we note that representation for the $\Xik$ in terms of scalar, vector, and tensor gauge transformation parameters is obtained by the replacement fields \rf{11032018-man01-07}-\rf{11032018-man01-15} with scalar, vector, and tensor gauge transformation parameters.

\noindent {\bf Solution to spin operators}. As we reviewed above, a problem for building  BRST-BV Lagrangian and gauge transformations amounts to a problem for building a realization for the spin operators and the $M^2$ which satisfy the (anti)commutation relations presented in \rf{15032018-man01-01}-\rf{15032018-man01-04}. In our approach, the continuous-spin field is described by $\Phik$ \rf{11032018-man01-01} which satisfies constraints
\rf{11032018-man01-02},\rf{11032018-man01-03}. This implies that we should find solution to (anti)commutation relations \rf{15032018-man01-01}-\rf{15032018-man01-04} which respects constraints \rf{11032018-man01-02},\rf{11032018-man01-03} . We find the following solution for the spin operators and the $M^2$:
\beq
\label{16082015-man01-08} &&  M^{\eta a} = \eta g_{\rho \upsilon}\alphab^a + A^a \gb_{\eta \upsilon}\etab\,,
\\
\label{16082015-man01-09} && M^{\eta\eta} = 2 \eta\etab\,,
\\
\label{16082015-man01-10} &&  M^\eta = \eta \upsilon l_{\rho \upsilon} +  \lb_{\eta \upsilon} \upsilonb \etab\,,
\\
\label{16082015-man01-11} && M^2 = m^2 \,,
\eeq
where the operator $A^a$ is defined in \rf{17032018-man01-06} and we use the notation
\beq
\label{07012018-10-a1} && g_{\rho \upsilon} = \Bigl(\frac{2N_\upsilon +d-4-2N_\rho}{2N_\upsilon +d-4}\Bigr)^{1/2} \,,
\\
\label{07012018-10-a2} && \gb_{\eta \upsilon} = - \Bigl(\frac{2N_\upsilon +d-4-2N_\eta}{2N_\upsilon +d-4}\Bigr)^{1/2} \,,
\\
\label{16082015-man01-17} && l_{\rho \upsilon} =  g_\upsilon \Bigl(\frac{2N_\upsilon +d-2-2N_\rho}{2N_\upsilon +d-2}\Bigr)^{-1/2} \,,
\\
\label{16082015-man01-18} && \lb_{\eta \upsilon} =  g_\upsilon \Bigl(\frac{2N_\upsilon + d-2-2N_\eta}{2N_\upsilon +d-2}\Bigr)^{-1/2}\,,
\\
\label{16082015-man01-19} && \hspace{1cm} g_\upsilon = \Bigl(\frac{1}{(N_\upsilon+1)(2N_\upsilon +d-2)} F_\upsilon \Bigr)^{1/2}\,,
\\
\label{16082015-man01-19-a1} && \hspace{1cm} F_\upsilon = \kappa^2 - N_\upsilon(N_\upsilon+d-3)m^2\,.
\eeq
In \rf{16082015-man01-11},\rf{16082015-man01-19-a1}, the $m$ is a mass parameter, while the $\kappa$ is a dimensionfull constant parameter.

Definition and detailed discussion of classical unitarity and irreducibility for continuous-spin field may be found in Sec.4. in Ref.\cite{Metsaev:2016lhs}. Briefly speaking, for the case under consideration, the classical unitarity and the irreducibility amount to the condition $F_\upsilon>0$ for all $N_\upsilon = 0,1,2,\ldots,\infty$. From \rf{16082015-man01-19-a1}, it is easy then to see that the classical unitarity and the irreducibility imply the following restrictions on the $\kappa$ and $m$: $\kappa^2>0$, $m^2 \leq 0$.

\noindent {\bf Interrelation between BRST-BV formulation and metric-like formulation}. We now establish the interrelation between our BRST--BV Lagrangian \rf{11032018-man01-16}
and Lagrangian obtained in the framework of  metric-like approach by using traceless
fields \rf{13032018-man01-16}. To this end we equate to zero all fields and
antifields having a nonzero ghost number,
\be \label{14032018-man01-01}
\ck = 0 \,, \qquad \cbk =  0\,, \qquad |\phi_{*\Ism}\rangle = 0 \,, \qquad  |c_*\rangle = 0 \,, \qquad |\phi_{*\IIsm}\rangle = 0 \,.
\ee
For the ghost numbers of fields and antifields, see relations \rf{17032018-man01-09},\rf{17032018-man01-10} in the Appendix. Making use of relations
\rf{14032018-man01-01}, it is easy to see that our BRST--BV Lagrangian
\rf{11032018-man01-16} takes the following form:
\be \label{14032018-man01-02}
\LL = \half \langle \phi_\Ism | (\Box - m^2)  |\phi_\Ism \rangle
- \half \langle \phi_\IIsm | (\Box - m^2) |\phi_\IIsm \rangle
- \langle \cb_*| \Lb_\Ism |\phi_\Ism\rangle
- \langle \cb_*| L_\IIsm |\phi_\IIsm\rangle
- \half \langle \cb_* | |\cb_*\rangle\,,
\ee
where the operators $\Lb_\Ism$, $L_\IIsm$ entering Lagrangian \rf{14032018-man01-02} are given in
\rf{13032018-man01-18},\rf{13032018-man01-19}. From Lagrangian \rf{14032018-man01-02}, using the Lagrangian  equation of motion for the antifield $|\cb_*\rangle$, we find the relation
\be \label{14032018-man01-03}
- |\cb_*\rangle =   \Lb_\Ism |\phi_\Ism\rangle  + | L_\IIsm |\phi_\IIsm\rangle\,,
\ee
which tells us that the antifield $|\cb_*\rangle$ is expressed in terms of the traceless fields $|\phi_\Ism\rangle$, $|\phi_\IIsm\rangle$. Plugging $|\cb_*\rangle$ \rf{14032018-man01-03} into \rf{14032018-man01-02}, we verify that BRST-BV Lagrangian \rf{14032018-man01-02} becomes the metric-like Lagrangian written in terms of traceless fields \rf{13032018-man01-16}.

\noindent {\bf Siegel gauge and global BRST and antiBRST transformations}. Making use of the Siegel gauge condition $|\phi_*\rangle = 0$, we find that the BRST-BV
Lagrangian \rf{11032018-man01-16}  leads to the simple
gauge-fixed Lagrangian given by
\be \label{14032018-man01-05}
\LL = \half \langle \phi_\Ism |(\Box - m^2)|\phi_\Ism \rangle
- \half \langle \phi_\IIsm |(\Box - m^2)|\phi_\IIsm \rangle
+ \langle \cb |(\Box - m^2) |c\rangle\,,
\ee
where, when passing from gauge invariant BRST-BV
Lagrangian \rf{11032018-man01-16} to gauge-fixed Lagrangian
\rf{14032018-man01-05}, we change the sign of the Faddeev--Popov
ghost field, $\cbk\rightarrow -\cbk$. One can make sure that the gauge-fixed Lagrangian given in
\rf{14032018-man01-05} is invariant under the global BRST and
antiBRST transformations,
\beq
\label{14032018-man01-06} && \hspace{-1cm} \ssf  |\phi_\Ism\rangle   =     G_\Ism\ck\,, \hspace{1cm} \ssf  |\phi_\IIsm\rangle   =     \Gb_\IIsm \ck\,, \qquad \ssf  \ck = 0\,, \hspace{1cm}  \ssf  \cbk   =  \Lb_\Ism |\phi_\Ism\rangle   + L_\IIsm |\phi_\IIsm\rangle  \,,\qquad
\\
\label{14032018-man01-07} && \hspace{-1cm} \ssfb  |\phi_\Ism\rangle  =    G_\Ism\ck\,,  \hspace{1cm} \ssfb  |\phi_\IIsm\rangle  =    \Gb_\IIsm  \cbk\,,\qquad   \ssfb   \cbk   = 0\,, \qquad \ssfb   \ck = -   \Lb_\Ism |\phi_\Ism\rangle - L_\IIsm |\phi_\IIsm\rangle\,,\qquad
\eeq
where the operators $\Lb_\Ism$, $L_\IIsm$ are given in \rf{13032018-man01-18},\rf{13032018-man01-19}, while the operators $G_\Ism$, $\Gb_\IIsm$ are
given in \rf{18032018-man01-01},\rf{18032018-man01-02}. Using the relation
\be
(\Lb_\Ism G_\Ism + L_\IIsm \Gb_\IIsm)\ck = (\Box - m^2)\ck
\ee
and the same relation for the $\cbk$, we verify that the global BRST and
antiBRST transformations given in \rf{14032018-man01-06},
\rf{14032018-man01-07} are nilpotent $\ssf^2=0$,
$\ssfb^2=0$, $\ssf\ssfb + \ssfb\ssf=0$ only for on-shell
Faddeev-Popov fields $\cbk$, $\ck$. Recent interesting discussion of (anti)BRST transformations
may be found in Refs.\cite{Gupta:2016oae}.

To summarize, we developed the BRST-BV formulation of the free massless and massive conti\-nuous-spin fields. As is well known the BRST approach turns out to be powerful method for the studying interaction vertices of higher-spin gauge field theories (see, e.g., Refs.\cite{Boulanger:2004rx}-\cite{Henneaux:2013gba}). We believe therefore that our BRST-BV formulation of the free continuous-spin fields will be helpful for the studying gauge invariant and Lorentz covariant interaction vertices of the continuous-spin field theory.

\bigskip
{\bf Acknowledgments}. This work was supported by the RFBR Grant No.17-02-00317.

\setcounter{section}{0}\setcounter{subsection}{0}
\appendix{ \large Notation and conventions  }

We use mostly positive flat metric $\eta^{ab}$, where vector indices of the $so(d-1,1)$ Lorentz algebra take values $a,b,c,e=0,1,\ldots ,d-1$. We drop the $\eta^{ab}$ in the scalar product:
$X^aY^a \equiv \eta_{ab}X^a Y^b$.

Derivatives of the space-time coordinates $x^a$ are denoted by $\partial^a \equiv \eta^{ab}\partial/\partial x^b$. Grassmann coordinate is denoted by $\theta$, $\theta^2=0$. The left derivative for the $\theta$ is denoted as $\partial_\theta = \partial/\partial \theta$. An integral over $\theta$ is normalized as $\int d\theta \theta =1$. Hermitian conjugation rules for the $x^a$, $\theta$ and their derivatives are defined as $(x^a,\theta)^\dagger = (x^a,\theta)$, $(\partial^a, \partial_\theta)^\dagger  = (-\partial^a, \partial_\theta)$.

Throughout this paper, creation operators $\alpha^a$, $\upsilon$, $\eta$, $\rho$ and the respective annihilation  operators $\alphab^a$, $\upsilonb$, $\rhob$, $\etab$ are referred to as oscillators.
(Anti)commutation relations, the vacuum $|0\rangle $ and the hermitian conjugation rules are defined by the relations
\beq
\label{17032018-man01-01} && [\alphab^a,\alpha^b] = \eta^{ab},    \ \quad \ [\upsilonb,\upsilon]=1, \ \quad \ \{\rhob,\eta\}=1\,, \ \quad \  \{\etab,\rho\} =1\,,\qquad
\\
\label{17032018-man01-02} && \alphab^a |0\rangle = 0\,,   \hspace{1.1cm} \upsilonb |0\rangle
 = 0\,,   \hspace{1cm} \etab |0\rangle = 0\,, \hspace{1cm} \rhob |0\rangle = 0\,,
\\
\label{17032018-man01-03} && \alpha^{a \dagger} = \alphab^a\,, \hspace{1.3cm} \upsilon^\dagger = \upsilonb\,, \hspace{1.3cm} \eta^\dagger = \etab\,, \hspace{1,4cm} \rho^\dagger = \rhob\,.
\eeq
We use the following shortcuts for products of the derivatives and the oscillators:
\beq
\label{17032018-man01-04} && \hspace{-1.9cm} \Box \equiv \partial^a \partial^a, \hspace{1cm} \alpar \equiv \alpha^a\partial^a  \hspace{1cm} \albpar\equiv \alphab^a\partial^a,  \hspace{1cm} \alpha^2 \equiv \alpha^a \alpha^a, \hspace{1cm} \alphab^2 \equiv  \alphab^a \alphab^a\,,
\\
\label{17032018-man01-05} && \hspace{-1.9cm} N_\alpha \equiv \alpha^a \alphab^a, \qquad    N_\upsilon \equiv \upsilon \upsilonb\,, \hspace{1cm}  N_\eta \equiv \eta \rhob\,, \qquad \ \ \ N_\rho \equiv \rho \etab\,,\qquad
\\
\label{17032018-man01-06} && \hspace{-1.9cm} A^a \equiv \alpha^a - \alpha^2\frac{1}{2N_\alpha  + d} \alphab^a\,,\hspace{1.1cm} \Pi^\smponetwo \equiv 1 - \alpha^2 \frac{1}{2(2N_\alpha+d)} \alphab^2\,.
\eeq
For a product of operators $A$, $B$, we use the hermitian conjugation rule $(AB)^\dagger = B^\dagger A^\dagger$.

The ghost numbers of the $\theta$, $\partial_\theta$, the oscillators, and the ket-vectors are given by the relations
\beq
\label{17032018-man01-07} &&     \gh(\alpha^a,\alphab^a,\upsilon,\upsilonb)=0\,,\qquad  \gh(\theta,\eta,\etab)=1,\qquad \gh(\partial_\theta,\rho,\rhob)=-1\,,
\\
\label{17032018-man01-09} && \gh(|\phi_\Ism\rangle)=0\,, \hspace{1cm} \gh(|c\rangle)=1\,, \hspace{1cm} \gh(|\cb\rangle)=-1\,, \hspace{1cm} \gh(|\phi_\IIsm\rangle)=0\,,
\\
\label{17032018-man01-10} && \gh(|\phi_{\Ism *}\rangle)=-1\,, \hspace{0.5cm} \gh(|c_*\rangle)=-2\,, \hspace{0.5cm} \gh(|\cb_*\rangle)=0\,,  \hspace{1.2cm} \gh(|\phi_{\IIsm *}\rangle) = -1\,.\qquad
\eeq
Using the notation $\XX$ for the $\theta$, $\partial_\theta$, and the oscillators, we note that ghost numbers \rf{17032018-man01-07} are fixed by the relation $[N_\FPsm^\intrm,\XX]= \gh(\XX)\XX$, where $N_\FPsm^\intrm = \theta\partial_\theta + N_\eta - N_\rho$. Ghost numbers of the ket-vectors \rf{17032018-man01-09},\rf{17032018-man01-10} are defined as eigenvalues of an external Faddeev-Popov operator denoted by $\Nbf_\FPsm^\ext$. Eigenvalues of  $\Nbf_\FPsm^\ext$ are fixed by the relation $(N_\FPsm^\intrm + \Nbf_\FPsm^\ext)\Phik=0$. In view of the relations $\gh(\alpha^a,\upsilon)=0$, $N_\FPsm^\intrm|0\rangle=0$, the ghost numbers of ket-vectors \rf{11032018-man01-07}-\rf{11032018-man01-15} coincide with ghost numbers of the corresponding fields on the right hand sides in \rf{11032018-man01-07}-\rf{11032018-man01-15}.
Ghost numbers of the gauge transformation parameters are found from the relation $(N_\FPsm^\intrm + \Nbf_\FPsm^\ext+1)\Xik=0$.

Hermitian conjugation rules we use for ket-vectors, bra-vectors, and tensor fields are as follows:
\beq
&& \Phibr = \Phik^\dagger\,,  \hspace{2.3cm} \phibr = \phik^\dagger\,, \hspace{2.3cm} \langle\phi_*| = |\phi_*\rangle^\dagger\,,
\\
&&  \langle\phi_{\Ism,\IIsm}| = |\phi_{\Ism,\IIsm}\rangle^\dagger\,, \qquad  \quad  \ \ \cbr = \ck^\dagger, \hspace{2.5cm} \cbbr = - \cbk^\dagger\,, \hspace{3cm}
\\
&& \langle\phi_{*\Ism,\IIsm}| = - |\phi_{*\Ism,\IIsm}\rangle^\dagger\,, \qquad  \langle c_*| = - |c_*\rangle^\dagger, \hspace{1.8cm}  \langle \cb_*| = |\cb_*\rangle^\dagger\,,
\\
&& \phi_{\Ism,\IIsm}^{a_1\ldots a_n\dagger} = \phi_{\Ism,\IIsm}^{a_1\ldots a_n},\hspace{1.3cm}  c^{a_1\ldots a_n\dagger} = c^{a_1\ldots a_n},\hspace{1.2cm}   \cb^{a_1\ldots a_n\dagger} = - \cb^{a_1\ldots a_n}\,,
\\
&& \phi_{*\Ism,\IIsm}^{a_1\ldots a_n\dagger} = - \phi_{\Ism,\IIsm}^{a_1\ldots a_n},\hspace{1cm}  c_*^{a_1\ldots a_n\dagger} = - c_*^{a_1\ldots a_n},\hspace{0.9cm}  \cb_*^{a_1\ldots a_n\dagger} = \cb_*^{a_1\ldots a_n}\,.
\eeq


\small

\begin{thebibliography}{30}



\parskip=-5pt


\bibitem{Bekaert:2006py}
  X.~Bekaert and N.~Boulanger,
 ``The Unitary representations of the Poincare group in any spacetime dimension,''
   in 2nd Modave Summer School in Theoretical Physics Modave, Belgium, August 6-12, 2006, 2006.
  hep-th/0611263.


\bibitem{Bekaert:2005in}
  X.~Bekaert and J.~Mourad,
  JHEP {\bf 0601}, 115 (2006)
  [hep-th/0509092].


\bibitem{Brink:2002zx}
  L.~Brink, A.~M.~Khan, P.~Ramond and X.~z.~Xiong,
  J.\ Math.\ Phys.\  {\bf 43}, 6279 (2002)
  [hep-th/0205145].



\bibitem{Savvidy:2003fx}
  G.~K.~Savvidy,
  Int.\ J.\ Mod.\ Phys.\ A {\bf 19}, 3171 (2004)
  [hep-th/0310085].
%
\\
%
  J.~Mourad,
  ``Continuous spin particles from a string theory,''
  hep-th/0504118.







\bibitem{Bengtsson:2013vra}
  A.~K.~H.~Bengtsson,
  JHEP {\bf 1310}, 108 (2013)
  [arXiv:1303.3799 [hep-th]].



\bibitem{Schuster:2014hca}
  P.~Schuster and N.~Toro,
  Phys.\ Rev.\ D {\bf 91}, 025023 (2015)
  [arXiv:1404.0675 [hep-th]].

\bibitem{Rivelles:2014fsa}
  V.~O.~Rivelles,
  Phys.\ Rev.\ D {\bf 91}, no. 12, 125035 (2015)
  [arXiv:1408.3576 [hep-th]].


\bibitem{Najafizadeh:2015uxa}
  X.Bekaert, M.Najafizadeh, M.R.Setare,
  Phys.\ Lett.\ B {\bf 760}, 320 (2016)
  [arXiv:1506.00973 [hep-th]].


\bibitem{Metsaev:2016lhs}
  R.~R.~Metsaev,
  Phys.\ Lett.\ B {\bf 767}, 458 (2017)
  [arXiv:1610.00657 [hep-th]].



\bibitem{Metsaev:2017ytk}
  R.~R.~Metsaev,
  Phys.\ Lett.\ B {\bf 773}, 135 (2017)
  [arXiv:1703.05780 [hep-th]].


\bibitem{Zinoviev:2017rnj}
  Y.~M.~Zinoviev,
  Universe {\bf 3}, no. 3, 63 (2017)
  [arXiv:1707.08832 [hep-th]].
%
\\
%
  D.~S.~Ponomarev and M.~A.~Vasiliev,
  Nucl.\ Phys.\ B {\bf 839}, 466 (2010)
  [arXiv:1001.0062 [hep-th]].



\bibitem{Najafizadeh:2017tin}
  M.~Najafizadeh,
  Phys.\ Rev.\ D {\bf 97}, no. 6, 065009 (2018)
  [arXiv:1708.00827 [hep-th]].

\bibitem{Bekaert:2017khg}
  X.~Bekaert and E.~D.~Skvortsov,
  Int.J.Mod.Phys.\ A {\bf 32}, no.23n24, 1730019 (2017)
  [arXiv:1708.01030].

\bibitem{Bekaert:2017xin}
  X.~Bekaert, J.~Mourad and M.~Najafizadeh,
  JHEP {\bf 1711}, 113 (2017)
  [arXiv:1710.05788 [hep-th]].


\bibitem{Metsaev:2017cuz}
  R.~R.~Metsaev,
  JHEP {\bf 1711}, 197 (2017)
  [arXiv:1709.08596 [hep-th]].

\bibitem{Khabarov:2017lth}
  M.~V.~Khabarov and Y.~M.~Zinoviev,
  Nucl.\ Phys.\ B {\bf 928}, 182 (2018)
  [arXiv:1711.08223 [hep-th]].

\bibitem{Metsaev:2017myp}
  R.~R.~Metsaev,
 ``Continuous-spin mixed-symmetry fields in AdS(5),''
  arXiv:1711.11007 [hep-th].

\bibitem{Alkalaev:2017hvj}
  K.~B.~Alkalaev and M.~A.~Grigoriev,
  JHEP {\bf 1803}, 030 (2018)
  [arXiv:1712.02317 [hep-th]].



\bibitem{Vasiliev:1990en}
  M.~A.~Vasiliev,
  Phys.\ Lett.\  B {\bf 243}, 378 (1990).
%
\\
%
  M.~A.~Vasiliev,
  Phys.\ Lett.\ B {\bf 285}, 225 (1992).
%
\\
%
  M.~A.~Vasiliev,
  Phys.\ Lett.\  B {\bf 567}, 139 (2003)
  [arXiv:hep-th/0304049].



\bibitem{Becchi:1974xu}
  C.~Becchi, A.~Rouet and R.~Stora,
  Phys.\ Lett.\ B {\bf 52}, 344 (1974).
%
\\
%
  I.~V.~Tyutin,
  ``Gauge Invariance in Field Theory and Statistical Physics in Operator Formalism,''  Lebedev Inst. preprint, No 39 (1975) [arXiv:0812.0580 [hep-th]].


\bibitem{Batalin:1981jr}
  I.~A.~Batalin and G.~A.~Vilkovisky,
  Phys.\ Lett.\ B {\bf 102}, 27 (1981).
%
\\
%
  I.~A.~Batalin and G.~A.~Vilkovisky,
  Phys.\ Rev.\ D {\bf 28}, 2567 (1983)
  [Phys.\ Rev.\ D {\bf 30}, 508 (1984)].



\bibitem{Metsaev:2015yyv}
  R.~R.~Metsaev,
  J.\ Phys.\ A {\bf 49}, no. 17, 175401 (2016)
  [arXiv:1511.01836 [hep-th]].



\bibitem{Metsaev:2015oza}
  R.~R.~Metsaev,
  Theor.\ Math.\ Phys.\  {\bf 187}, no. 2, 730 (2016)
  [arXiv:1508.07928 [hep-th]].



\bibitem{Metsaev:2014vda}
  R.~R.~Metsaev,
  Theor.\ Math.\ Phys.\  {\bf 181}, no. 3, 1548 (2014)
  [arXiv:1407.2601 [hep-th]].




\bibitem{Buchbinder:2001bs}
  I.L.Buchbinder, A.Pashnev, M.Tsulaia,
  Phys.\ Lett.\  B {\bf 523}, 338 (2001)
  [arXiv:hep-th/0109067].
%
\\
%
  I.~L.~Buchbinder, V.~A.~Krykhtin, P.~M.~Lavrov,
  Nucl.\ Phys.\  B {\bf 762}, 344 (2007)
  hep-th/0608005



\bibitem{Sagnotti:2003qa}
  A.~Sagnotti and M.~Tsulaia,
  Nucl.\ Phys.\  B {\bf 682}, 83 (2004)
  [arXiv:hep-th/0311257].
%
\\
%
  A.~Fotopoulos and M.~Tsulaia,
  Int.\ J.\ Mod.\ Phys.\  A {\bf 24}, 1 (2009)
  [arXiv:0805.1346 [hep-th]].



\bibitem{Alkalaev:2009vm}
  K.~B.~Alkalaev and M.~Grigoriev,
  Nucl.\ Phys.\ B {\bf 835}, 197 (2010)
  [arXiv:0910.2690 [hep-th]].
%
\\
%
  K.~Alkalaev and M.~Grigoriev,
  Nucl.\ Phys.\ B {\bf 853}, 663 (2011)
  [arXiv:1105.6111 [hep-th]].


\bibitem{Reshetnyak:2010ga}
  A.~A.~Reshetnyak,
  Phys.\ Part.\ Nucl.\  {\bf 41}, 976 (2010)
  [arXiv:1002.0124 [hep-th]].
%
\\
%
  I.~L.~Buchbinder and A.~Reshetnyak,
  Nucl.\ Phys.\ B {\bf 862}, 270 (2012)
  [arXiv:1110.5044 [hep-th]].



\bibitem{Reshetnyak:2018fvd}
  A.~Reshetnyak,
  ``Constrained BRST- BFV Lagrangian formulations for Higher Spin Fields in Minkowski Spaces,''
  arXiv:1803.04678 [hep-th].
%
  ``Constrained BRST-BFV and BRST-BV Lagrangians for half-integer HS fields on $R^{1,d-1}$,''
  arXiv:1803.05173 [hep-th].

\bibitem{Siegel:1984wx}
  W.~Siegel,
  Phys.\ Lett.\ B {\bf 149}, 157 (1984)
  [Phys.\ Lett.\ B {\bf 151}, 391 (1985)].
%
\\
%
  W.~Siegel,
  ``Fields,''
  hep-th/9912205.


\bibitem{Francia:2002aa}
  D.~Francia and A.~Sagnotti,
  Phys.\ Lett.\ B {\bf 543}, 303 (2002)
  [hep-th/0207002].
%
\\
%
  D.~Francia and A.~Sagnotti,
  Phys.\ Lett.\ B {\bf 624}, 93 (2005)
  [hep-th/0507144].




\bibitem{Gupta:2016oae}
  S.~Gupta and R.~Kumar,
  Int.\ J.\ Mod.\ Phys.\ A {\bf 31}, no. 34, 1650173 (2016)
  [arXiv:1608.01613 [hep-th]].
%
\\
%
  S.~Gupta and R.~Kumar,
  Int.\ J.\ Theor.\ Phys.\  {\bf 55}, no. 2, 927 (2016)
  [arXiv:1411.6357 [hep-th]].




\bibitem{Boulanger:2004rx}
  N.~Boulanger and S.~Cnockaert,
  JHEP {\bf 0403}, 031 (2004)
  [hep-th/0402180].
%
\\
  X.~Bekaert, N.~Boulanger and S.~Cnockaert,
  J.\ Math.\ Phys.\  {\bf 46}, 012303 (2005)
  [hep-th/0407102].

\bibitem{Fotopoulos:2010ay}
  A.~Fotopoulos and M.~Tsulaia,
  JHEP {\bf 1011}, 086 (2010)
  [arXiv:1009.0727 [hep-th]].


\bibitem{Metsaev:2012uy}
  R.~R.~Metsaev,
  Phys.\ Lett.\ B {\bf 720}, 237 (2013)  [arXiv:1205.3131 [hep-th]].



\bibitem{Henneaux:2013gba}
  M.Henneaux, G.L.Gomez and R.Rahman,
  JHEP {\bf 1401}, 087 (2014)  [arXiv:1310.5152 [hep-th]].
%
\\
%
  M.Henneaux, G.L.Gomez and R.Rahman,
  JHEP {\bf 1208}, 093 (2012)
  [arXiv:1206.1048 [hep-th]].
%
\\
%
  P.~Dempster and M.~Tsulaia,
  Nucl.\ Phys.\ B {\bf 865}, 353 (2012)  [arXiv:1203.5597 [hep-th]].
%
\\
%
  M.~Taronna,
  JHEP {\bf 1204}, 029 (2012)  [arXiv:1107.5843 [hep-th]].
%
\\
%
  I.Buchbinder, P.Dempster, M.Tsulaia,
  Nucl.Phys.B {\bf 877}, 260 (2013)  [arXiv:1308.5539 [hep-th]].



\end{thebibliography}
\end{document}